\documentclass[%
reprint,
superscriptaddress,
amsmath,amssymb,
aps,
]{revtex4-1}
\usepackage{graphicx}% Include figure files
\usepackage{dcolumn}% Align table columns on decimal point
\usepackage{bm}% bold math
\usepackage{xcolor}
\usepackage{multirow}
\usepackage{amsmath,amssymb,amsfonts}
\usepackage{array,tabularx}
\usepackage{booktabs}                 % 三线表- 短细横线
\usepackage[colorlinks,linkcolor=blue, urlcolor=blue, anchorcolor=blue, citecolor=blue]{hyperref}
\usepackage{hyperref}
\begin{document}
\maxdeadcycles=1000
\preprint{APS}

\title{Towards a nuclear isomer quantum battery}% Nuclear Isomer Quantum Batteries Towards a laser?driven nuclear isomer quantum battery Force line breaks with \\
%\thanks{A footnote to the article title}%

\author{Ying-Bo Gao}
\affiliation{College of Physics and Electronic Engineering, Northwest Normal University, Lanzhou, 730070, China}
\author{Fu-Quan Dou}
\email{doufq@nwnu.edu.cn}
\affiliation{College of Physics and Electronic Engineering, Northwest Normal University, Lanzhou, 730070, China}
\affiliation{Gansu Provincial Research Center for Basic Disciplines of Quantum Physics, Lanzhou, 730000, China}
%\date{\today}% It is always \today, today,
             %  but any date may be explicitly specified
%文章中关于原子核的单词的校对
\begin{abstract}
Quantum batteries (QBs)---quantum devices governed by the principles of quantum mechanics---hold great promise for next-generation energy storage. However, most existing research efforts focus on atomic or molecular systems featuring low-energy and short-lived energy levels, leaving the exploitation of high-energy, ultra-stable nuclear energy levels for energy storage as a critical unaddressed challenge. Here, we propose an innovative design of referred to as nuclear isomer quantum batteries (NIQBs), whose energy storage unit is typically composed of two-level and three-level nuclei incorporating nuclear isomers. The charging dynamics is driven by the interaction between the nuclear system and x-ray free electron laser (XFEL). Compared to QBs based on atomic systems, NIQBs deliver substantial performance enhancements, with stored energy and average charging power enhanced by factors of $10^{1}$--$10^{6}$ and $10^{6}$--$10^{11}$, respectively, and a markedly extended lifetime range spanning from microseconds to $10^5$ years. Notably, the vast majority of NIQBs enable complete energy extraction as their excited-state lifetimes exceed the laser-nuclear interaction time, rendering spontaneous emission negligible. The proposed NIQBs framework is compatible with diverse nuclear systems, enabling tailored nucleus selection for varied operating conditions. Our results provide a feasible pathway toward realizing high-performance QBs with superior energy-storage efficiency.

\end{abstract}

%\keywords{Suggested keywords}%Use showkeys class option if keyword
                              %display desired
\maketitle

%\tableofcontents

\emph{Introduction}---The escalating global energy demand and intensifying global energy crisis has spurred ongoing efforts to innovate energy storage techniques \cite{iea_weo_2025}. Concurrently, rapid advances in quantum technology and growing miniaturization of electronic devices have motivated the application of quantum-mechanical principles to energy storage \cite{quach20232195}. In 2013, Alicki and Fannes first proposed the concept of quantum batteries (QBs)---quantum systems engineered for energy storage and extraction \cite{PhysRevE.87.042123}.
Leveraging the quantum resources, QBs hold the potential to achieve more efficient work extraction and faster charging processes \cite{RevModPhys.96.031001,Ferraro2026-op}.
Considerable efforts have been directed towards constructing QB models \cite{PhysRevE.100.032107,PhysRevA.97.022106,PhysRevLett.120.117702,PhysRevLett.124.130601,PhysRevLett.125.236402,Dou_2020,PhysRevLett.127.100601,PhysRevResearch.4.013172,PhysRevB.105.115405,
PhysRevLett.120.117702,PhysRevE.104.044116,PhysRevA.109.032201,PhysRevE.104.024129,PhysRevE.101.062114,
PhysRevA.106.022618,PhysRevA.109.062432,PhysRevB.100.115142,PhysRevLett.122.047702,PhysRevLett.132.210402} and analysis of QB performance \cite{quach2020,PhysRevLett.118.150601,PhysRevLett.122.210601,PhysRevLett.125.040601,PhysRevLett.129.130602,PhysRevLett.131.030402,PhysRevLett.132.240401,PhysRevLett.131.060402,PhysRevLett.132.210402,PhysRevA.106.032212,PhysRevLett.128.140501,PhysRevA.107.022215,
PhysRevB.109.235432,PhysRevLett.134.220402,kzvn-dj7v,2026-0106}.
Despite substantial progress, two critical challenges remain in QBs field, as most models utilize atomic or molecular energy levels: low stored energy stemming from their intrinsic energy levels, and self-discharging (or aging) with degradation arising from short excited-state lifetimes and decoherence \cite{PhysRevA.100.043833,PhysRevA.107.023725,PhysRevLett.132.090401}. Consequently, further exploration of QBs with high energy and high efficiency is vital and urgent for advancing energy storage technologies.

Fortunately, the advent of x-ray free electron lasers (XFELs) \cite{RevModPhys.84.1177,RevModPhys.88.015006,RevModPhys.69.1085,HUANG2021100097} has enabled nuclear coherent population transfer (NCPT) at transition energies of several hundred keV, which is implemented through a variety of quantum control techniques \cite{RevModPhys.84.1177,PhysRevLett.96.142501},
ranging from $\pi$-pulse methods \cite{Shore01051991,Amiri2023}, stimulated raman adiabatic passage (STIRAP) \cite{VITANOV200155,RevModPhys.89.015006,LIAO2011134,PhysRevC.87.054609,MANSOURZADEHASHKANI2021122119}, mixed-state inverse engineering (MIE) \cite{PhysRevApplied.16.044028,PhysRevC.110.034606} to coincident pulses \cite{PhysRevA.85.043407,PhysRevC.94.054601,PhysRevC.96.044619}. More interestingly, as a result of nuclear structure effects, certain nuclei can possess nuclear isomers \cite{Aprahamian2005}, which are generally prepared via radiative transitions from higher excited states \cite{PhysRevC.87.054609,PhysRevC.105.064313}. Compared with ordinary nuclear excited states, these isomeric states exhibit markedly longer lifetimes and extremely narrow linewidths, thereby enabling the storage of substantial amounts of energy over extended time scales \cite{Walker1999,doi:10.1126/science.1080552,LIAO2011134,Liao2014}. %machine learning \cite{gdpq-k6lj}

Notably, nuclear systems inherently exhibit long-lifetime, high-energy two-, three-, or multi-level configurations. These unique characteristics render nuclear systems promising candidates for next-generation energy storage devices. %Prior work has demonstrated that,
 Analogous to atomic or molecular-scale two-level systems employed as work qubits, atomic nuclei also fulfill the basic criteria for qubits implementation \cite{Vagizov2014,Pla2013-qy,Metzler_2023,Ladd2010,Morton2008}. %A natural question arises: Might atomic nuclei serve as the work cells for QBs-which could then be referred to as nuclear quantum batteries (NQBs)? %boyles2014gamma,
 This encourages researchers to propose using atomic nuclei as work cell for QBs---which could then be referred to as nuclear isomer quantum batteries (NIQBs). However, capitalizing on these nuclear advantages---particularly exploiting the properties of nuclear isomeric states for QB applications and realizing efficient energy storage and extraction---remains an unresolved challenge. % However, leveraging these nuclear merits-particularly the properties of isomeric states for QBs and achieving efficient energy storage/extraction remain unresolved.
 Against this backdrop, two key questions naturally emerge: How can nuclear systems---featuring long-lived, high-energy-level structures---be harnessed to develop viable energy storage devices, and can such devices enable efficient energy storage and extraction?

In this Letter, we propose an innovative design of NIQBs, which involves coupling of nuclear systems with an XFEL to enhance the energy storage capability of QBs. We employ two- and three-level nuclear systems with isomeric state as the battery cells, and investigate the stored energy, average charging power, and work extraction efficiency of NIQBs. We find that the NIQBs have extremely superior charging performance: large energy, high average power, long lifetime, and full extractability of stored energy with the nuclear system remaining in a pure state.

\emph{Model}---We consider an NIQB modeled as a $d$-level nuclear system coupled to external laser field as depicted in Fig. \ref{fig1}. The NIQB system can be described by the following Hamiltonian
\begin{eqnarray}\label{H}
H(t)=H_{0}+H_{1}(t).
\end{eqnarray}
Here, the Hamiltonian of the battery is given by $H_{0}=\sum_{n=1}^{d}\varepsilon_{n}| n \rangle\langle n |$~($d=2, 3$ for two- and three-levels system, respectively),
where $\varepsilon_{n}$ ($|n\rangle$) is the $n$th eigenenergies (eigenstate).
The time-dependent Hamiltonian $H_{1}$ serves as the quantum charger and describes the interaction between the battery and the external field, which generally depends on the structure of the system \cite{PhysRevLett.96.142501,PhysRevA.84.053429}.
We consider a fully coherent XFEL source of pump and Stokes lasers \cite{PhysRevC.87.054609}, to drive population transfer in nuclear systems.
In the two-level system, the states $|1\rangle$ and $|2\rangle$ are coupled by pump laser with Rabi frequency $\Omega_{p}$.
In the three-level system, the states $|1\rangle$ and $|2\rangle$ are coupled by pump laser with Rabi frequency $\Omega_{p}$, while $|2\rangle$ and $|3\rangle$ states are coupled by the Stokes laser with Rabi frequency $\Omega_{s}$.
The Rabi frequency can be represented as \cite{PhysRevC.94.054601,PhysRevC.110.034606}
\begin{eqnarray}\label{Omega}
\Omega_{p(s)}\!=\!\Omega_{0p(s)}\sqrt{I_{p(s)}}\exp\!\left\{\!-\!\left[\!\frac{\gamma(1\!+\!\beta)(t\!-\!\tau_{p(s)})}{\sqrt{2}T_{p(s)}} \!\right]^{2}\! \right\}.
\end{eqnarray}
The $\Omega_{0p(s)}$ correlates with the nuclear characteristic parameters, further details and numerical values of the parameters are provided in the Supplementary Material (SM) \cite{SupplementalMaterial}.
$\tau_{p(s)}$ and $I_{p(s)}$ are temporal peak position and effective peak intensity of pump and Stokes laser pulse, respectively.
The nuclear beam moves with velocity $v=\beta c$ and relativistic factor $\gamma=1/\sqrt{1-\beta^{2}}$, where $\beta$ is the velocity of the nuclear particle and $c$ represents velocity of light in vacuum.
\begin{figure}[tbp]
    \centering
    \includegraphics[width=0.485\textwidth]{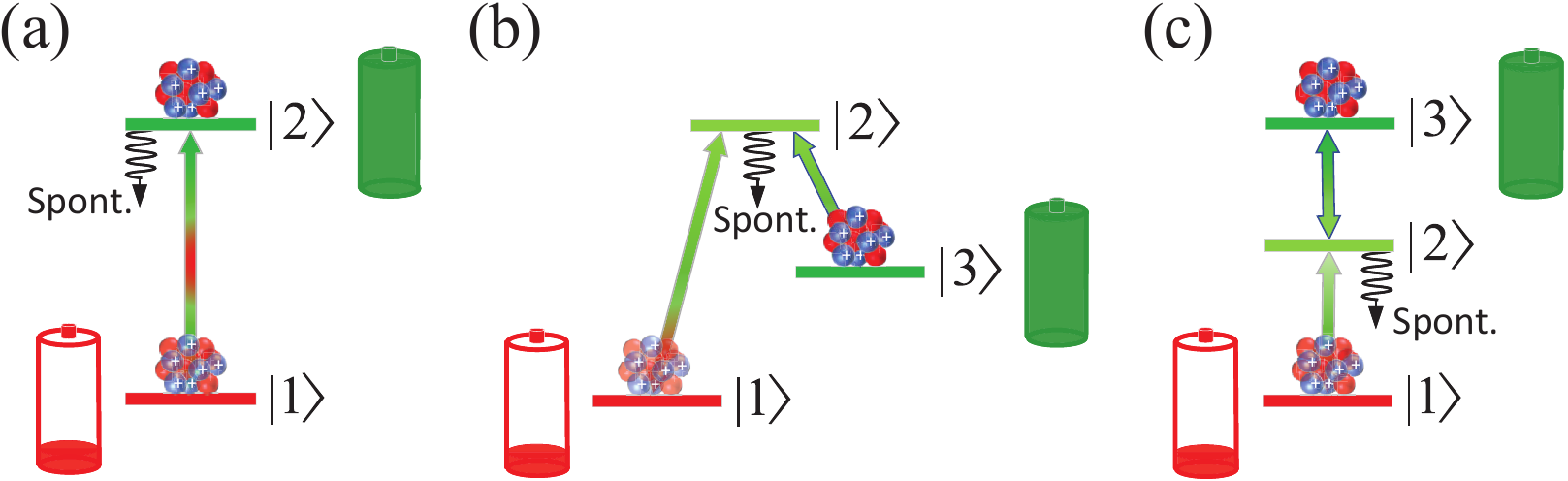}%width=0.50\textwidth
    \caption{Schematic of the NIQBs setup. (a) Schematic of a two level nuclear system, composed of the ground state $|1\rangle$ and an excited state $|2\rangle$ (isomeric state) of the nucleus. All population in $|1\rangle$ denotes an empty battery, while complete population in $|2\rangle$ corresponds to a fully charged state. (b) Schematic of nuclear three-level $\Lambda$ and (c) ladder-type schemes, consisting of the nuclear ground state $|1\rangle$ and two different energy levels of excited states $|2\rangle$ and $|3\rangle$ (isomeric state). All population in $|1\rangle$ means an empty battery, and a fully charged state is achieved with complete population in $|3\rangle$.}
    \label{fig1}
\end{figure}

We model the dynamics of the system via the density matrix approach. Assuming that the battery is initially in the ground state $\rho\left(0\right)=|1\rangle\langle 1|$, the dynamics is driven by the interaction Hamiltonian $H_{1}\left(t\right)$.
In the interaction picture, the nuclear system dynamics is governed by master equation for the density matrix $\rho_{\text{int}}(t)$ \cite{PhysRevC.87.054609,PhysRevC.94.054601,PhysRevC.110.034606}
\begin{eqnarray}\label{rho(t)}
\frac{\partial}{\partial t}\rho_{\text{int}}(t)=\frac{1}{i\hbar}[H_{\text{int}}(t),\rho_{\text{int}}(t)]+\rho_{s}(t),
\end{eqnarray}
where $H_{\text{int}}(t)=U^{\dagger}\left(t\right) H_{1}\left(t\right) U\left(t\right)$, and $\rho_{\text{int}}(t)=U^{\dagger}\left(t\right) \rho\left(t\right) U\left(t\right)$, with $U\left(t\right)=e^{-iH_{0}t / \hbar}$.
The decoherence matrix $\rho_{s}(t)$ is caused by the spontaneous emission of excited states.

The stored energy and the average charging power are crucial figure of merits to evaluate the performance of NIQB. The stored energy $E(t)$ is
\begin{eqnarray}\label{E(t)}
E(t)=\text{Tr}\left[H_{0}\rho_{\text{int}}(t)\right]-\text{Tr}\left[H_{0}\rho_{\text{int}}(0)\right].
\end{eqnarray}
The average charging powers $P(t)$ is defined as
\begin{eqnarray}\label{P(t)}
P(t)=\frac{E(t)}{t}.
\end{eqnarray}

\emph{Two-level NIQB}---The primary approach involves coherent population transfer in a two-level NIQB via $\pi$-pulse \cite{Amiri2023}, enabling controlled energy storage. In this scheme, the pulse area of the Rabi frequency must be regulated to satisfy the condition at the end of time evolution, given by $A=\int_{-\infty}^{+\infty} \Omega(t)\,dt=\pi$. Under the resonance condition $ \varepsilon_{2}-\varepsilon_{1}=\gamma (1+\beta) \hbar \omega_{p}$, the interaction Hamiltonian of the two-level NIQB can be expressed as
\begin{eqnarray}\label{H(I2)}
H_{int}(t)=\frac{\hbar}{2} \begin{pmatrix}0 & \Omega_{p} \\ \Omega_{p} & 0 \end{pmatrix},
\end{eqnarray}
where the pump pulse with a Rabi frequency of $\Omega_{p}$, and the decoherence matrix is
\begin{eqnarray}\label{rho(s2)}
\rho_{s}=\frac{\Gamma}{2} \begin{pmatrix} 2\rho_{22} & -\rho_{12} \\ -\rho_{21} & -2\rho_{22} \end{pmatrix},
\end{eqnarray}
here $\Gamma$ is linewidth of excited state $|2\rangle$.

%To characterize the performance of a two-level NIQB, it is essential to evaluate both the stored energy and the charging power.
Typically, the ground state and a metastable state of a nucleus are selected as the two working levels for NIQBs. For special nuclear species---where transitions occur between the ground state and the isomeric state---the nuclear isomeric state is directly adopted as the metastable state \cite{vonderWense2020}. Consequently, a nucleus's ground state and isomeric state serve as the fundamental cells of NIQBs. Fortunately, a number of naturally occurring atomic nuclei exhibit exclusive transitions between the ground state and their isomeric states. We take the $^{193}\text{Ir}$, $^{117}\text{Sn}$ and $^{113}\text{Cd}$ nuclei as representative examples. All exhibit stable ground states---enabling the avoidance of extraneous radiation---with natural abundances of $62.7\%$ ($^{193}\text{Ir}$), $7.68\%$ ($^{117}\text{Sn}$) and $12.227\%$ ($^{113}\text{Cd}$), and isomeric state half-lives of $10.53$ d (days), $14$ d, and $14.1$ y (years), respectively \cite{PhysRevC.36.1132,UNTERWEGER2002125,Flynn01081965}.
Their specific level structures, characteristic and performance parameters are provided in the SM \cite{SupplementalMaterial}. These NIQBs are efficiently charged by employing a seeded XFEL \cite{FELDHAUS1997341,SALDIN2001357} with a photon energy of $12.4\,\text{keV}$, a bandwidth of $10\,\text{meV}$, and a pulse duration of $0.1\,\text{ps}$. The time evolution of the Rabi frequencies, the stored energy and the average charging power are depicted in Figs.~\ref{fig2}(a)--\ref{fig2}(c). To further analyze the charging mechanism, we also calculate the population dynamics of the nuclear systems, and the time-dependent population of the target state is shown Fig.~\ref{fig2}(d). It is clear that NIQBs can attain considerable stored energy on the order of $10^{5}\,\text{eV}$ ($^{117}\text{Sn}$ is $3.15\times 10^5$ eV) and maintain stable storage over a certain period of time (corresponding to the moment when the population of the target state is completely transferred), whereas QBs based on atomic systems reach only the order of $10^{-2}-10^{0}\,\text{eV}$. Since the interaction time between the nucleus and the laser lies on the femtosecond scale, the maximum charging power of the NIQB reaches the order of watts. In contrast, previous QBs exhibit charging powers only at the $10^{-13}$--$10^{-10}$ watt level.
\begin{figure}[hbp]
    \centering
    \includegraphics[width=0.485\textwidth]{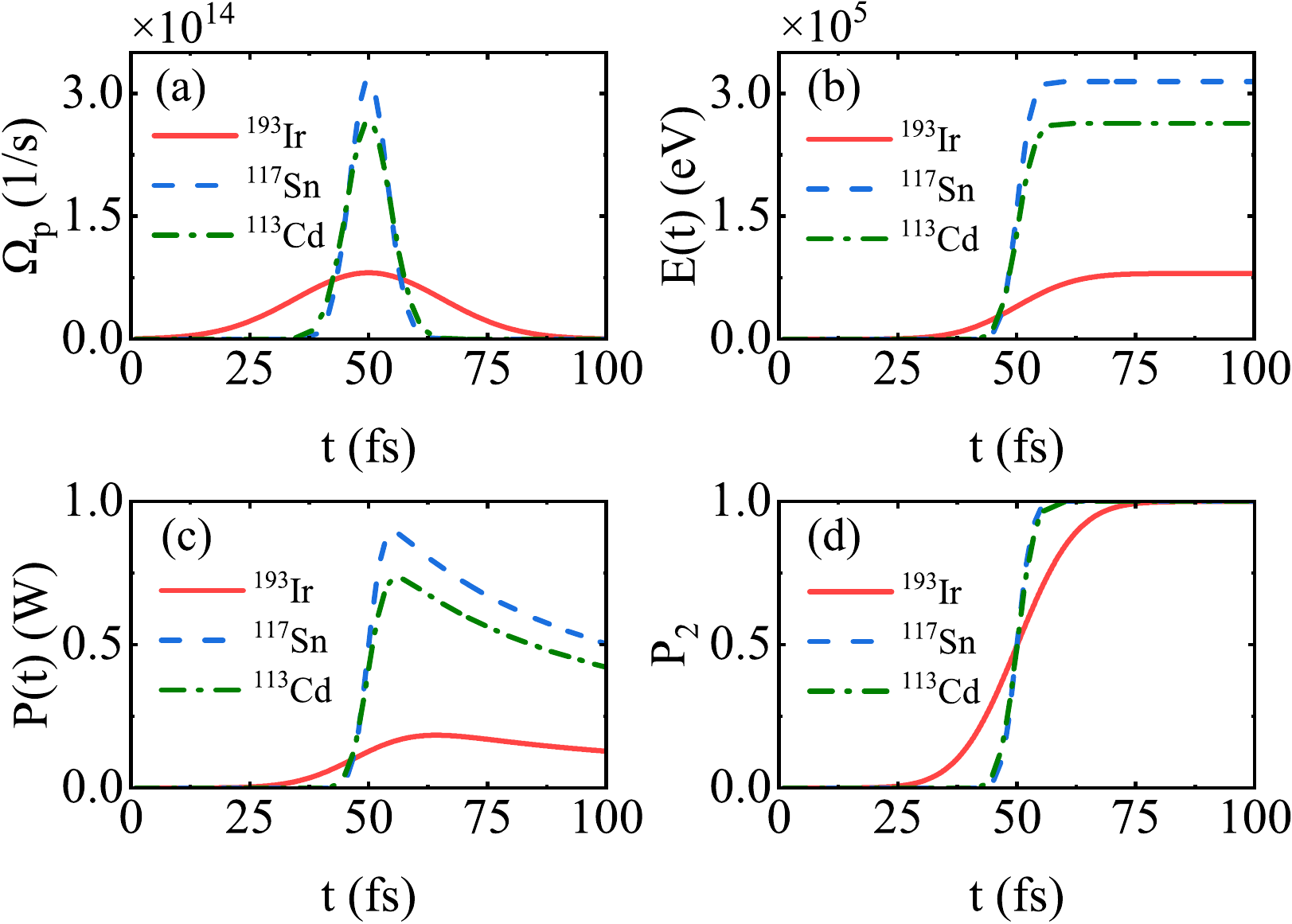}
    \caption{(a) The pulse shape controlled by $\pi$-pulse, (b) the time evolution of the stored energy $E(t)$ (in units of $eV$), (c) the average charging power $P(t)$ (in units of $W$), and (d) the population of the target state for three selected two-level nuclear systems: $^{193}$Ir,$^{117}$Sn, and $^{113}$Cd.}
    \label{fig2}
\end{figure}

\emph{Three-level NIQB}---For the three-level NIQB, direct transitions from the ground state to the isomeric state are generally forbidden; population of the isomeric state is thus typically achieved via an intermediate energy level, corresponding to two canonical configurations: the $\Lambda$-type system (Fig. \ref{fig1} (b)) and the ladder-type system (Fig. \ref{fig1} (c)). This stepwise process allows for precise quantum control via STIRAP. The interaction Hamiltonian matrix for the three-level NIQB satisfying the one-photon resonance condition $\varepsilon_{3(2)}-\varepsilon_{1}=\gamma(1+\beta)\hbar \omega_{p(s)}$ can be written as
\begin{eqnarray}\label{H(I3)}
H_{int}(t)=\frac{\hbar}{2} \begin{pmatrix}0 & \Omega_{p} & 0\\ \Omega_{p} & 0 & \Omega_{s}\\ 0 & \Omega_{s} & 0\end{pmatrix}.
\end{eqnarray}
%with $\Omega_{p}$ and $\Omega_{s}$ being the Rabi frequencies of pump pulse and Stokes pulse, respectively.
 We focus on the battery charging process governed by adiabatic dynamics, where the NIQB evolves along the dark-state eigenstate \cite{RevModPhys.77.633}. The decoherence matrix is
\begin{eqnarray}\label{rho(s31)}
\rho_{s}=\frac{\Gamma}{2} \begin{pmatrix} 2B_{21}\rho_{22} & -\rho_{12} & 0 \\ -\rho_{21} & -2\rho_{22} & -\rho_{23} \\ 0 & -\rho_{32} & 2B_{23}\rho_{22}\end{pmatrix}
\end{eqnarray}
for $\Lambda$-type, and
\begin{eqnarray}\label{rho(s32)}
\rho_{s}=\frac{\Gamma}{2} \begin{pmatrix} 2\rho_{22} & -\rho_{12} & 0 \\ -\rho_{21} & -2\rho_{22} & -\rho_{23} \\ 0 & -\rho_{32} & 0\end{pmatrix}
\end{eqnarray}
for ladder-type, where $\Gamma$ is the linewidth of excited state $|2\rangle$ and is also spontaneous decay rate of $|2\rangle$. $B_{2i}$ is branching ratio of $|2\rangle \rightarrow |i\rangle (i=1,3)$ spontaneous decay.
\begin{figure*}[htbp]
    \centering
    \includegraphics[width=0.985\textwidth]{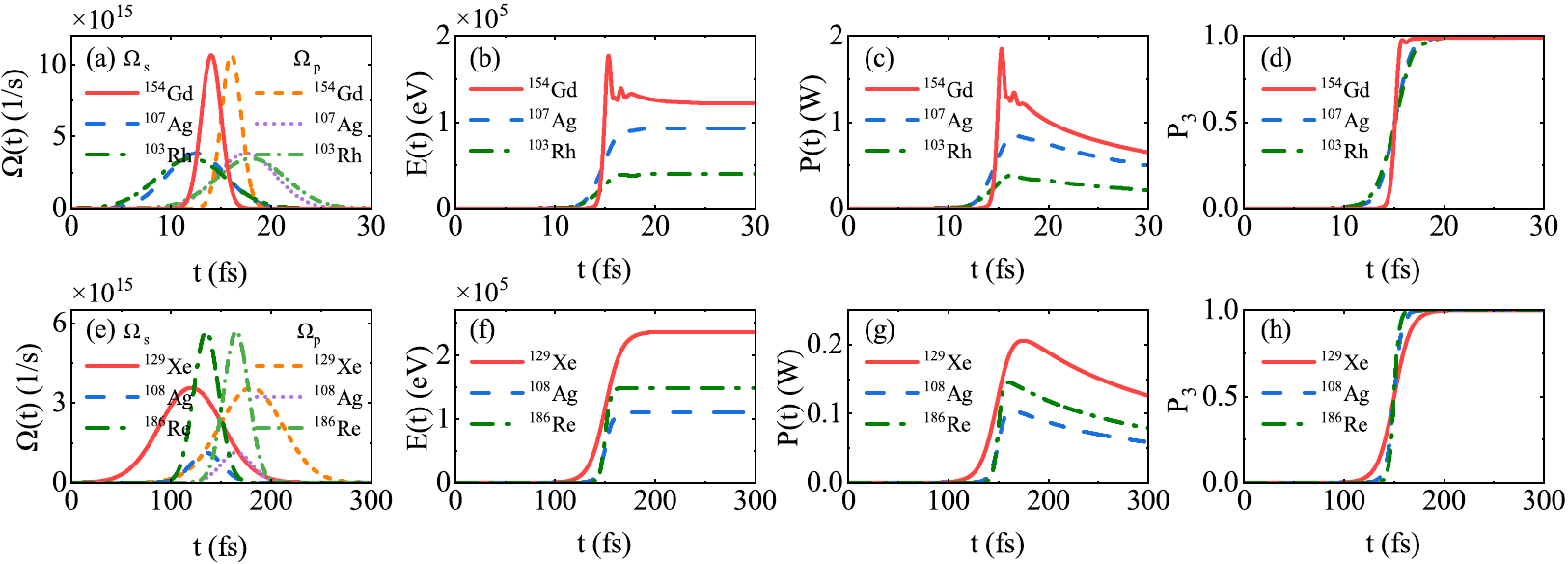}
    \caption{(a), (e) The pulse shape controlled by STIRAP, (b), (f) time evolution of the stored energy $E(t)$ (in units of $eV$), (c), (g) the average charging power $P(t)$ (in units of $W$) and (d), (h) the population $P_3$ of the target state $|3\rangle$ for selected three-level nuclear systems: $\Lambda$-type of $^{154}$Gd, $^{107}$Ag and $^{103}$Rh (a--d) and ladder-type of $^{129}$Xe, $^{108}$Ag and $^{186}$Re (e--h), respectively.}
    \label{fig3}
\end{figure*}

In our scheme, the STIRAP protocol, employing a counterintuitive pulse sequence as shown in Fig.~\ref{fig3}(a) and (e), realizes complete and stable charging of the NIQB through full population transfer from the initial state $ |1\rangle$ via the intermediate excited state $|2\rangle $ to the target state $|3\rangle$. In particular, the target state $|3\rangle$ is typically chosen as a nuclear isomer \cite{Aprahamian2005}, whose long lifetime makes it well suited for energy storage. Owing to the dark state nature of this protocol, the state $|2\rangle $ remains almost unpopulated during the charging process, thereby minimizing the influence of spontaneous emission.
The time evolution of the stored energy, the average charging power and the population of target state are presented in Figs.~\ref{fig3}(b)--\ref{fig3}(d) and ~\ref{fig3}(f)--\ref{fig3}(h).
Similar to the two-level NIQB, the three-level NIQB can achieve a stored energy on the order of $10^{5}$~eV and the average charging power on the order of watts. This enhanced energy storage capability originates from the distinctive properties of nuclear isomers, in which the combination of nuclear structure effects inhibits its decay, resulting in a lifetime significantly longer than that of ordinary nuclear excited states \cite{Aprahamian2005}. For $\Lambda$-type $^{154}$Gd, $^{107}$Ag and $^{103}$Rh NIQBs, the lifetime span from nanoseconds to hours. Compared with $\Lambda$-type counterparts, ladder-type NIQBs exhibit a higher stored energy and a much longer lifetime (from hours to years). For instance, a single $^{108}$Ag NIQB has a stored energy of $109.45$ keV with a lifetime of several hundred years, whereas $^{186}$Re has a lifetime of $10^5$ years. Further analysis demonstrates excellent laser intensity robustness for the charging performance of three-level NIQBs (see SM \cite{SupplementalMaterial} for details). %All their specific level structures and characteristic parameters are provided in the SM \cite{SupplementalMaterial}. % %Their specific level structures and characteristic parameters are provided in the Supplemental Material (SM) \cite{SupplementalMaterial}

\emph{The extractable work}---According to the second law of thermodynamics, complete extraction of the stored energy from a quantum battery is generally unattainable. For this reason, ergotropy $W(t)$ is introduced as the canonical measure of the maximum extractable work in quantum systems \cite{Allahverdyan2004}. Meanwhile, the ratio of ergotropy to stored energy serves as another practical figure of merit for characterizing NIQB performance, defined as
\begin{eqnarray}\label{R(t)}
R(t)=\frac{W(t)}{E(t)}.
\end{eqnarray}
In addition, the purity of the NIQB is also considered, and its corresponding measure is given by
\begin{eqnarray}\label{mP(t)}
\mathcal{P}(t)=\text{Tr}[\rho_{int}(t)^{2}].
\end{eqnarray}

To analyze the practically extractable energy, we calculate the ratio $R(t)$ and the purity $\mathcal{P}(t)$ for NIQBs with different level configurations (the corresponding figures are provided in SM \cite{SupplementalMaterial}). Our results show a consistent trend between system purity and the energy extraction ratio of NIQBs: higher purity yields a higher energy extraction ratio. % in Fig.~\ref{fig4}.
Consistent with QB \cite{PhysRevLett.122.047702}, when NIQB system is pure, the ergotropy coincides with the stored energy, i.e., $W(t)=E(t)$, indicating that the stored energy can be completely extracted. The two-level NIQB remains essentially in a pure state throughout the entire evolution. Although spontaneous emission from the excited state generally induces decoherence, this effect is negligible in these NIQBs since the lifetime of the excited state is much longer than the interaction time between the laser and the nucleus. Specifically, the isomeric states in $^{193}\text{Ir}$, $^{117}\text{Sn}$, and $^{113}\text{Cd}$ feature lifetimes ranging from days to years, whereas the interaction time is on the femtosecond timescale. Consequently, the stored energy can be extracted almost completely. In three-level $\Lambda$-type systems, these intermediate levels possess higher energy than the target isomeric states, whereas in ladder-type configurations, the energy of the intermediate levels is lower than that of the isomeric states. For the $^{154}\text{Gd}$ nucleus with an excited state lifetime of $1.54$~fs, which is shorter than the interaction time between the laser and the nucleus, the system does not remain in a pure state during the energy transfer process, resulting in incomplete extraction of the stored energy. In contrast, for other nuclear systems, the excited-state lifetimes exceed the laser-nuclear interaction time, so that these systems exhibit behavior analogous to that of the two-level system and the stored energy can thus be nearly fully extracted. Our analysis further identifies a fundamental relationship in NIQBs: higher system purity allows the ergotropy to approach the stored energy more closely.

\emph{Discussions and Conclusions}---Unlike traditional nuclear batteries (or isotope batteries), which convert energy from radioactive decay (in the form of $\alpha, \beta, \gamma$ rays or decay-generated heat) into electricity for long-lived power supplies \cite{olsen2012betavoltaic,PRELAS2014117,Spencer2019,Li2024}, our NIQBs store energy in the high-level isomeric states of atomic nuclei. Therefore, NIQBs can also serve as the front-end basis for traditional nuclear batteries. For simplicity, we only consider several atomic nuclei as examples in the above analysis. In fact, one can utilize other atomic nuclei and isomers to construct NIQBs as required. Tables S3-S4 in SM \cite{SupplementalMaterial} present some typical nuclear systems with distinct physical characteristics and summarize their key performance metrics for two- and three-levels NIQBs, encompassing a diverse set of nuclear isomers with large energy and broad lifetime distributions. Two-level NIQBs ($^{133}\text{Xe}$, $^{193}\text{Ir}$, $^{117}\text{Sn}$, $^{127}\text{Te}$, $^{113}\text{Cd}$, and $^{93}\text{Nb}$) exhibit mean lifetimes on the order of days and stored energy spanning from a few to hundreds of keV. Three-level NIQBs, by comparison, feature nuclear isomers with distinctive energy and lifetime traits for targeted quantum storage applications. For instance, the $^{229}$Th isomer has a uniquely low nuclear excitation energy ($8.355$ eV), well within the reach of current laser technology \cite{Kraemer2023,PhysRevLett.132.182501,PhysRevLett.133.013201,PhysRevLett.127.052501,PhysRevLett.133.223001,Ooi2026}, making it well-suited for compact energy storage. The $^{63}\text{Ni}$ isomer has also short lifetime with $1.69\mu$s. However, it possesses high stored energy ($87.23$ keV). Others $\Lambda$-type NIQBs systems ($^{107}\text{Ag}$, $^{144}\text{Pr}$, $^{103}\text{Rh}$, $^{189}\text{Os}$, $^{152}\text{Eu}$, and $^{121}\text{Sn}$) exhibit mean lifetimes spanning from seconds to decades and stored energy from a few to tens of keV. Ladder-type NIQBs ($^{195}\text{Pt}$, $^{129}\text{Xe}$, $^{121}\text{Te}$, $^{119}\text{Sn}$, $^{108}\text{Ag}$, and $^{186}\text{Re}$), by contrast, exhibit mean lifetimes spanning well beyond one day---extending to $2\times10^5$ years---and stored energy all exceeding $100$ keV. All other parameter values 
\begin{figure}[t]
    \centering
    \includegraphics[width=0.485\textwidth]{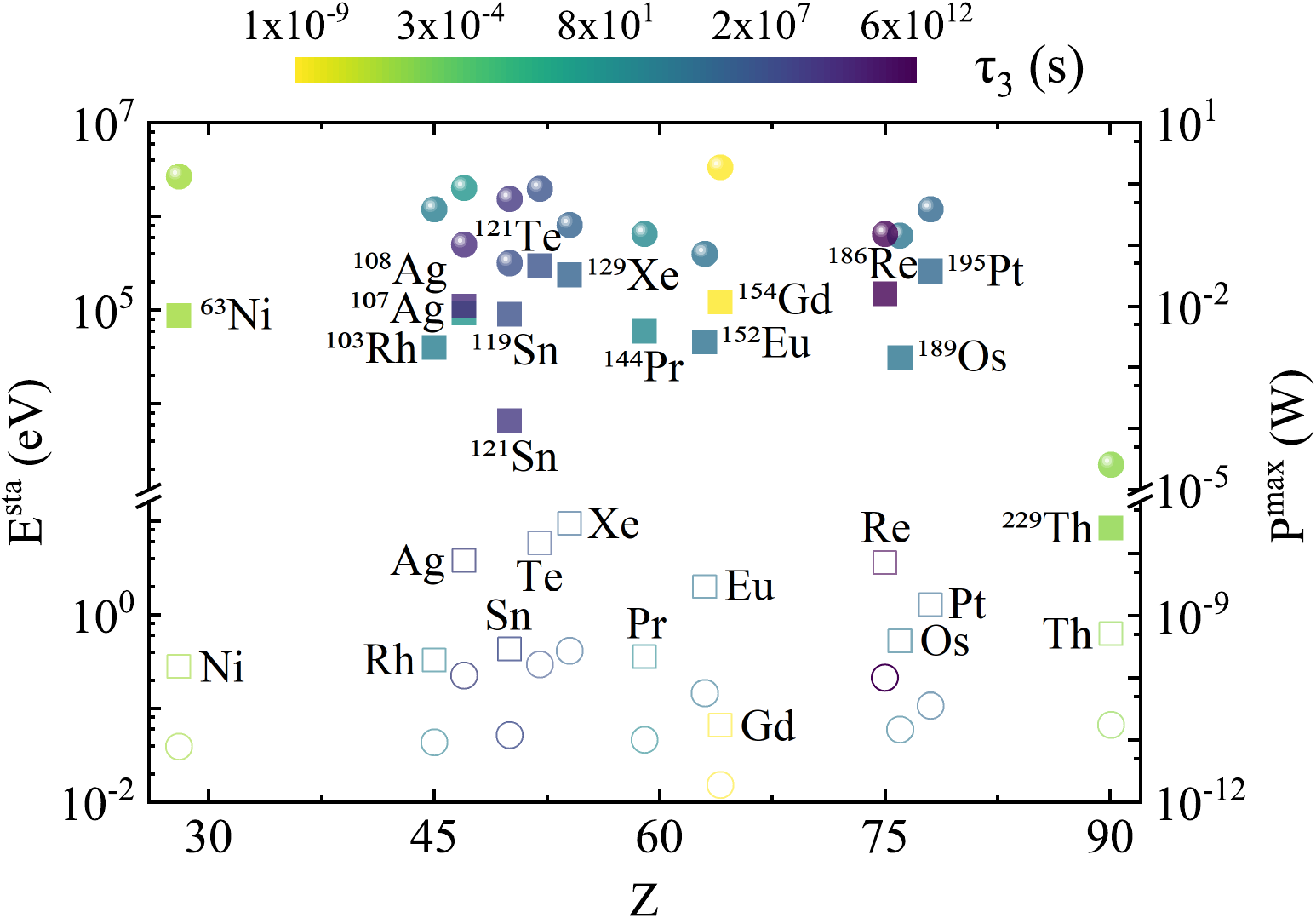}
    \caption{The stable stored energy (in units of $eV$) and the maximum average charging power (in units of $W$) of three-level NIQBs and QBs for typical nuclei and the corresponding atom, respectively. $Z$ denotes the neutron number of a nucleus. Solid squares and solid circles correspond to the stored energy and average charging power of NIQBs, respectively; hollow squares and hollow circles denote the same parameters for QBs. Colors indicate the lifetimes of the nuclear isomeric states.}
    \label{fig4}
\end{figure}
are available in Ref.~\cite{NNDC2025}. These examples demonstrate the versatility and scalability of NIQBs across a wide range of application scenarios. We also calculated the stored energy and average charging power of the three-level atomic QBs (see Table S5 in SM \cite{SupplementalMaterial}) corresponding to these typical nuclear species, with the comparison results presented in Fig. \ref{fig4}. Evidently, the NIQBs exhibit a dramatic performance advantage over their atomic QBs counterparts in terms of stored energy, average charging power and lifetime: specifically, their stored energy and charging power are enhanced by approximately $10^1-10^6$-fold and $10^{6}-10^{11}$-fold, with the smallest enhancement observed in Th and the largest in Gd. With the exception of the Th case, the NIQBs achieve a maximum stored energy of several hundred keV, with maximum average charging power reaching the watt-scale level.

In summary, we have proposed a general theoretical framework for an NIQB composed of a nuclear ensemble coupled to an XFEL and characterized its performance via stored energy, average charging power, and ergotropy as key metrics. We have demonstrated that employing a nuclear system with high-energy and long-lifetime isomeric state configuration as an energy storage cell confers enhanced energy storage capability and charging power during femtosecond-scale interactions, yielding a stored energy on the order of ~$10^5$~eV, an average charging power of ~$1$~W and the lifetimes ranging from nanoseconds to $10^5$ years.
 We have investigated energy extraction in NIQBs and found a consistent correlation between the energy extraction ratio and system purity; most two- and three-levels NIQBs enable complete energy extraction by virtue of maintaining a pure quantum state. The performance of NIQBs may be further optimized via various coherent nuclear population transfer techniques, such as shortcut to adiabaticity \cite{PhysRevLett.105.123003,RevModPhys.91.045001,Dou2021} and recent machine learning \cite{RevModPhys.91.045002,RevModPhys.94.031003,gdpq-k6lj,Sun_2025,PhysRevLett.133.243602}. The prospective advantages of NIQBs include an extraordinarily high energy density, which may surpass that of many existing energy storage technologies by orders of magnitude, along with superior charging power and exceptional anti-aging performance. Our results offer a feasible pathway for high-efficient energy storage, enriching a theoretical configurations for designing next-generation energy storage devices.

\emph{Acknowledgments}---The work is supported by the National Natural Science Foundation of China (Grants No. 12475026) and the Natural Science Foundation of Gansu Province (No. 25JRRA799).

\emph{Data availability}---The data that support the findings of this article are not publicly available. The data are available from the authors upon reasonable request.

\bibliography{reference}
{
%%%%%%%%%%%%%%%% End Matter %%%%%%%%%%%%%%%%%
\appendix*
\renewcommand{\theequation}{A\arabic{equation}}
%\renewcommand{\theequation}{\Alpha{A}}
%\arabic{equation}
\onecolumngrid

\begin{center}
	\textbf{End Matter}
\end{center}

\twocolumngrid
\setcounter{equation}{0}
%This End Matter provides that the proof of the stored energy in the interaction picture and the eigenvalues and eigenvector of three-level QB system.

\emph{Proof of Eq. (\ref{E(t)})}---The stored energy $E(t)$ is defined as the energy difference between the final and initial states of the battery: %$E(t)=\text{Tr}[H_{0}\rho(t)]-\text{Tr}[H_{0}\rho(0)].$
\begin{eqnarray}
E(t)=\text{Tr}[H_{0}\rho(t)]-\text{Tr}[H_{0}\rho(0)].
\end{eqnarray}
In the interaction picture, the population of each energy level can be obtained as $P_{n}=\text{Tr}[ \hat{P}_{n} \rho_{\text{int}}\left(t\right)]=\text{Tr}[ \hat{P}_{n} \rho\left(t\right)]$, where $\hat{P}_{n}$ is the operator $\hat{P}_{n}=|n\rangle \langle n|$. Therefore, $\text{Tr}[H_{0} \rho_{\text{int}}\left(t\right)]=\text{Tr}[H_{0} \rho\left(t\right)]$ and the stored energy reads
\begin{eqnarray}
E(t)=\text{Tr}[H_{0}\rho_{\text{int}}(t)]-\text{Tr}[H_{0}\rho_{\text{int}}(0)].
\end{eqnarray}

\emph{The eigenvalues and eigenstates of the three-level NIQB}---For a three-level NIQB, the time-dependent eigenstates $|\lambda_{n}(t) \rangle (n=0,+,-)$ are
\begin{eqnarray}
|\lambda_{0}(t) \rangle&=&\cos{\theta}|1\rangle - \sin{\theta}|3\rangle, \nonumber \\
|\lambda_{\pm}(t) \rangle&=&\frac{1}{\sqrt{2}} \sin{\theta} |1\rangle\!\pm\!\frac{1}{\sqrt{2}} |2\rangle \!+\!\frac{1}{\sqrt{2}} \cos{\theta} |3\rangle,
\end{eqnarray}
with the associated eigenvalues given by $\lambda_{0}=0$ and $\lambda_{\pm}=\pm \hbar\Omega(t)/2$, where $\Omega^{2}(t)=\Omega^{2}_{p}(t) + \Omega^{2}_{s}(t)$. The mixing angle $\theta$ is defined by $\tan{\theta(t)}=\Omega_{p}(t)/\Omega_{s}(t)$. Specifically, the eigenstate $|\lambda_{0}\rangle$ corresponds to the well-known dark state, a coherent superposition of the initial state $|1\rangle$ and the target state $|3\rangle$.

\emph{Definition of ergotropy}---The ergotropy $W(t)$ is defined as the maximum amount of energy that can be extracted from the quantum system \cite{Allahverdyan2004}:
%\vspace{-5pt}
%\vskip -7mm
\begin{eqnarray}\label{W(t)}
W(t)=\text{Tr}[H_{0}\rho_{int}(t)]-\text{Tr}[H_{0}\tilde{\rho}_{int}(t)],
\end{eqnarray}
where $\tilde{\rho}_{int}(t)=\sum_{n}r_{n}|n \rangle \langle n|$ is the passive state, $r_{n}$ are the eigenvalues of $\rho_{int}(t)$ arranged in descending order, and $|n\rangle$ are the eigenstates of $H_{0}$ with the corresponding eigenvalues $\varepsilon_{n}$ sorted in ascending order.
}
\onecolumngrid
\newpage

\begin{center}
    \Large\textbf{Supplemental Material for ``Towards a nuclear isomer quantum battery''}
\end{center}

%\title{Supplemental Material for ``The Ergotropy in Quantum Batteries''}
\begin{center}
Ying-Bo Gao\textsuperscript{1}, and Fu-Quan Dou\textsuperscript{1,2,*}\\
\vspace{5pt}
\small
\textsuperscript{1}\emph{College of Physics and Electronic Engineering, Northwest Normal University, Lanzhou, 730070, China}\\
\textsuperscript{2}\emph{Gansu Provincial Research Center for Basic Disciplines of Quantum Physics, Lanzhou, 730000, China}
\end{center}

\vspace{25pt}
% 重置计数器，并且加上S 前缀
\setcounter{section}{0}
\renewcommand{\thesection}{\Roman{section}}
\setcounter{figure}{0}
\renewcommand{\thefigure}{S\arabic{figure}}
\renewcommand{\theHfigure}{S\arabic{figure}}   % 新增
\setcounter{equation}{0}
\renewcommand{\theequation}{S\arabic{equation}}
\setcounter{table}{0}
\renewcommand{\thetable}{S\arabic{table}}
%\title{Supplemental Material for ``Towards a nuclear isomer quantum battery''}
%\author{Ying-Bo Gao}
%\affiliation{College of Physics and Electronic Engineering, Northwest Normal University, Lanzhou, 730070, China}
%
%\author{Fu-Quan Dou}
%\email{doufq@nwnu.edu.cn}
%\affiliation{College of Physics and Electronic Engineering, Northwest Normal University, Lanzhou, 730070, China}
%\affiliation{Gansu Provincial Research Center for Basic Disciplines of Quantum Physics, Lanzhou, 730000, China}

%\maxdeadcycles=1000
%\preprint{APS}

%\section{PROOF OF PROPOSITION 2}
\noindent
  %we provide additional details and data supporting the main text.This Supplemental Material elaborates on the relevant details of the research work and provides the specific parameters required for the calculations. Section. \ref{section1} presents the nuclear level structures. We analyzes the charging robustness of the three-level nuclear quantum battery (NQB) in Sec. ~\ref{section2}. Characteristic parameters of the relevant nuclei are summarized in Sec.~\ref{section3}. Charging-related parameters for different nuclear species employed as storage units in two- and three-level systems are summarized in Sec.~\ref{section4}.
  This Supplemental Material provides additional details on the specific nuclear parameters, laser intensity robustness, the time evolution of the energy extraction ratio and purity, and charging performance for nuclear isomer quantum batteries (NIQBs) in the main text.  Section~\ref{section1} presents schematic energy-level structures of the considered nuclei. The robustness of the charging process on laser intensity in three-level NIQBs, the time evolution of the energy extraction ratio and purity are discussed in Sec.~\ref{section2}. Section~\ref{section3} displays characteristic parameters of the relevant nuclei, while charging performance and laser parameters for different two- and three-level NIQBs systems are given in Sec.~\ref{section4}.
\vspace{30pt}
\maketitle

\section{Nuclear level structure}\label{section1}
\begin{figure}[htbp]
    \centering
    \includegraphics[width=0.95\textwidth]{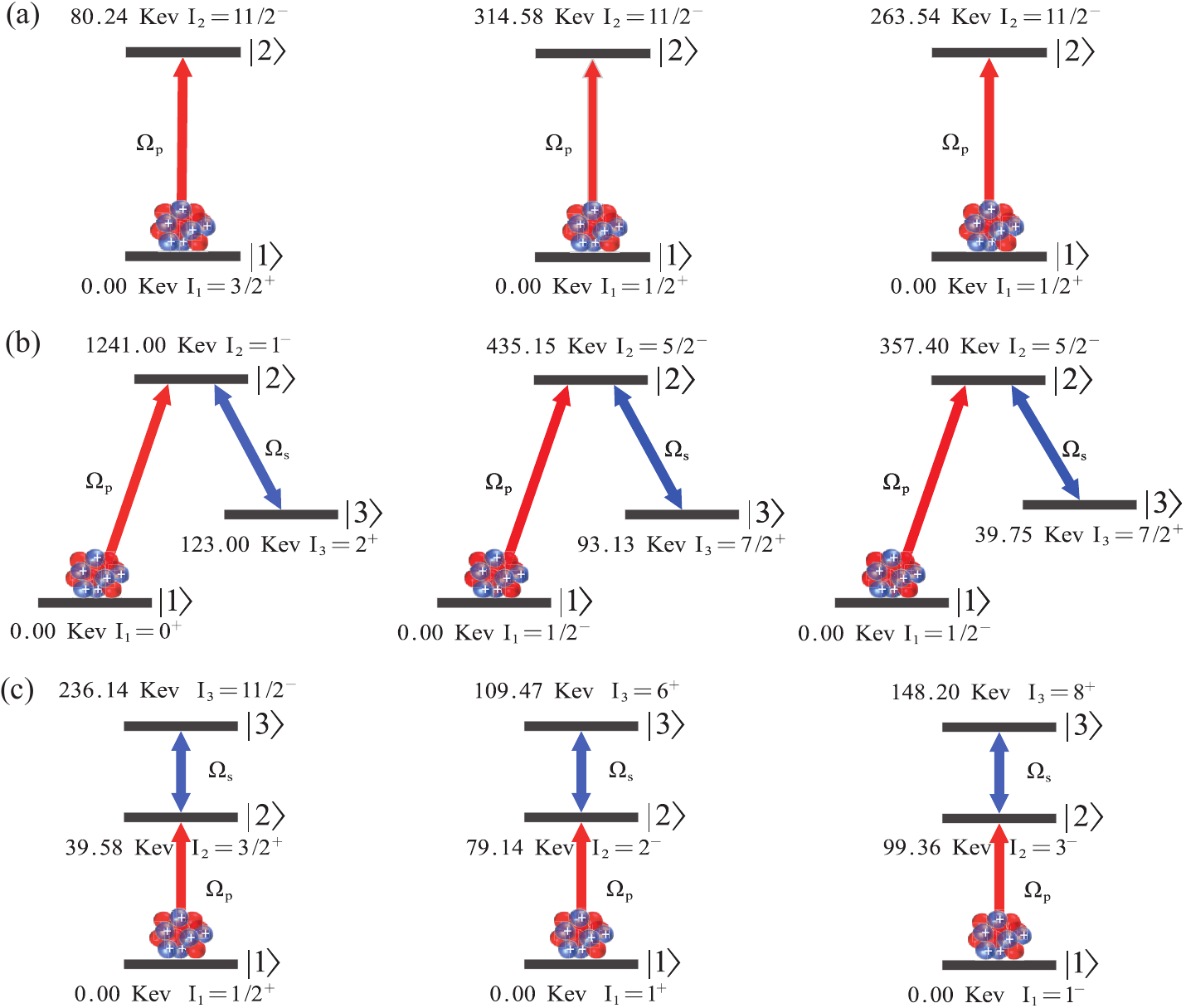}%width=0.50\textwidth
    \caption{(a) Two-level nuclear configurations for $^{193}\text{Ir}$, $^{117}\text{Sn}$ and $^{113}\text{Cd}$, formed by the ground and isomeric states, respectively. %Each state is characterized by its energy, angular momentum, and parity. $\Omega_p$ denotes the Rabi frequency of the coupling between two levels.
     (b) Three-level $\Lambda$-type nuclear configurations formed by the ground, excited, and isomeric states: $^{154}\text{Gd}$, $^{107}\text{Ag}$, and $^{103}\text{Rh}$; (c) Three-level ladder-type configurations in $^{129}\text{Xe}$, $^{108}\text{Ag}$, and $^{186}\text{Re}$.  Each state is labeled with its energy, angular momentum, and parity. $\Omega_p$ and $\Omega_s$ correspond to the Rabi frequency coupling between the respective levels.}
    \label{figs1}
\end{figure}

Here, we present detailed nuclear level structures for representative two- and three-level nuclear QB configurations. The specific energy-level structures of the considered nuclei are shown in Fig.~\ref{figs1}.

For two-level nuclear isomer quantum batteries (NIQBs) systems, nuclei including $^{193}\text{Ir}$, $^{117}\text{Sn}$, and $^{113}\text{Cd}$ are specifically selected for their ability to form a well-defined two-level configuration--consisting of a stable ground state and a long-lived nuclear isomeric state--along with their relatively high natural abundances \cite{PhysRevC.36.1132,PhysRevC.59.2836,yzlp-wcyf}. %Alexeev2019
 This two-level architecture, anchored by the stable ground state and the long-lived isomeric state, suppresses additional radiative transitions originating from spontaneous decay, thereby facilitating coherent quantum manipulation and stable energy storage--two core requirements for NIQB operation. Furthermore, their relatively high natural abundances enhance the experimental accessibility of these two-level NIQB systems, lowering barriers for practical implementation.

For three-level NIQB systems, $\Lambda$-type configurations in $^{154}\text{Gd}$, $^{107}\text{Ag}$, and $^{103}\text{Rh}$ similarly feature stable ground states and relatively high natural abundances \cite{PhysRev.82.486,PhysRevLett.113.032501}, enabling the adoption of a more complex three-level architecture while preserving overall stability and experimental feasibility. In contrast, the selection of ladder-type NIQB configurations for $^{129}\text{Xe}$, $^{108}\text{Ag}$, and $^{186}\text{Re}$ is primarily governed by the half-lives of their constituent nuclear isomeric states \cite{PhysRevC.93.064315,PhysRevC.30.2026,PhysRevC.92.054304}. The long half-lives of these isomers facilitate effective energy storage over extended time scales, thus endowing these ladder-type configurations with inherent advantages for long-term quantum energy storage.
\section{Laser Intensity robustness for Three-Level NIQBs and the time evolution of the energy extraction ratio and purity}\label{section2}
By varying the peak intensities of the pump and Stokes lasers over a prescribed range, we quantitatively evaluated the stored energy and the energy extraction efficiency, thereby enabling a further characterization of the charging performance advantages of our three-level NIQBs.
\begin{figure}[htbp]
    \centering
    \includegraphics[width=0.985\textwidth]{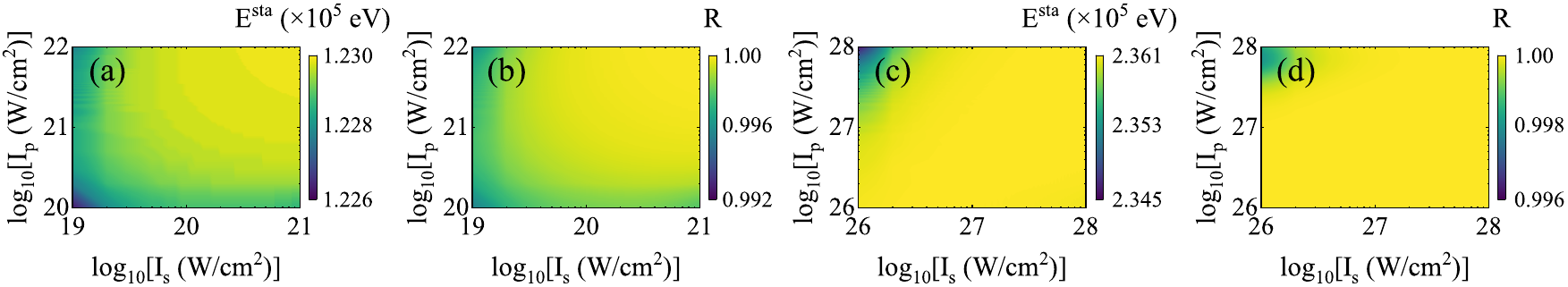}%width=0.50\textwidth
    \caption{Contour plots of the stored energy $E$ (in units of eV) and the extractable energy ratio $R$ as functions of the pump laser intensity $I_{p}$ and the Stokes laser intensity $I_{s}$: (a), (b) for the $\Lambda$-type three-level NIQB system $^{154}\text{Gd}$, and (c), (d) for the ladder-type three-level NIQB system $^{129}\text{Xe}$.}
    \label{figs2}
\end{figure}
%The dependence of the stored energy and ratio of the NIQB on the pump peak intensity $I_{p}$ and Stokes peak intensity $I_{s}$ is illustrated in Fig.~\ref{figs2}.
The variations of the stored energy and energy extraction efficiency of the NIQB with respect to the pump peak intensity $I_p$ and the Stokes peak intensity $I_s$ are illustrated in Fig.~\ref{figs2}. Taking $^{154}\text{Gd}$ and $^{129}\text{Xe}$ as representative examples, the stored energy is consistently maintained on the order of $10^{5}\text{eV}$ over the considered range of laser intensities. Crucially, we find that increasing the laser intensity significantly enhances the adiabatic effect, thereby effectively suppressing the decoherence of the excited state $|2\rangle$ during the charging process. This suppression directly boosts the energy extraction efficiency and ultimately ensuring that the battery yields the maximum amount of valuable work within the charging time.

\begin{figure}[htbp]
    \centering
    \includegraphics[width=0.985\textwidth]{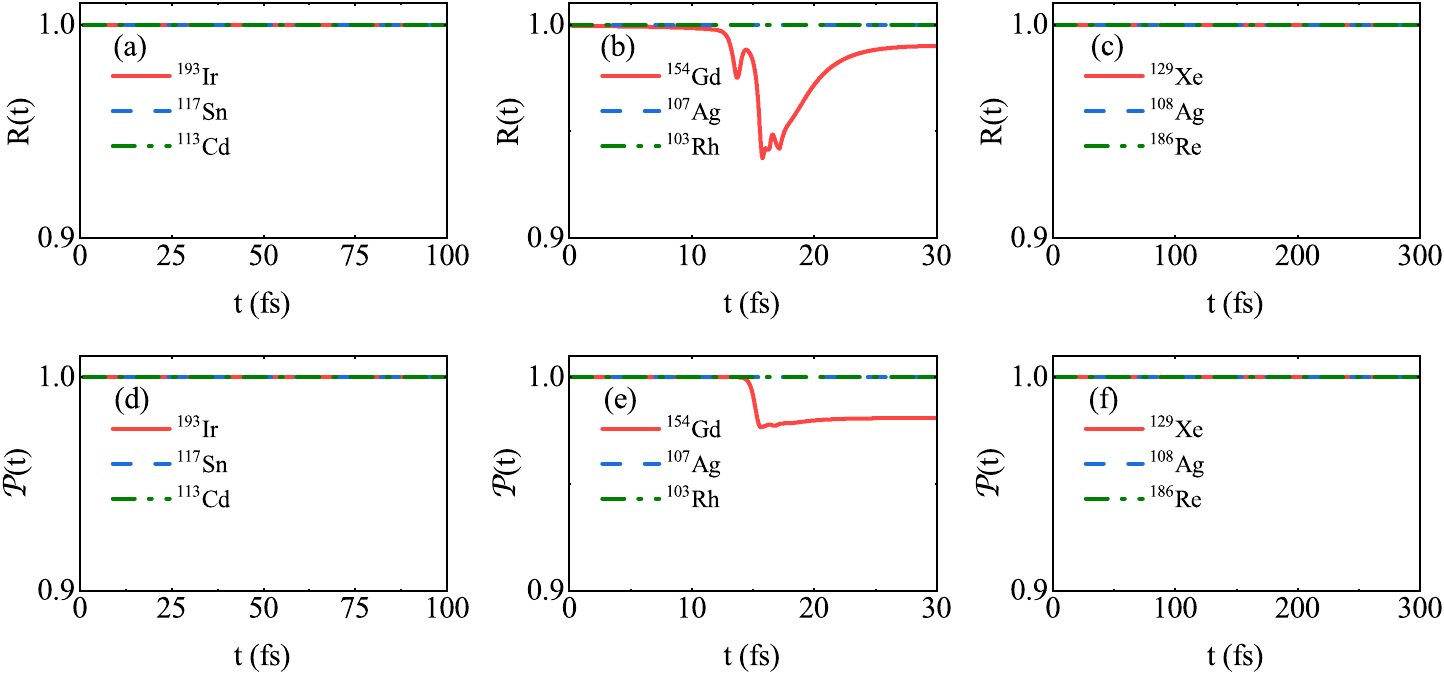}
    \caption{The energy extraction ratio $R(t)$ and purity $ \mathcal{P}(t)$ as a function of $t$ for (a) the two-level NIQBs, (b) the three-level $\Lambda$-type NIQBs, and (c) the three-level ladder-type NIQBs.}
    \label{figs3}
\end{figure}
Figure~\ref{figs3} illustrates the time evolution of the energy extraction ratio $R(t)$ and purity $ \mathcal{P}(t)$. For two-level and three-level ladder-type NIQBs, full energy extraction is achievable: the lifetimes of the isomeric state far exceed the interaction times, thereby rendering decoherence negligible, and system purity confirms the preservation of a pure quantum state. In contrast, for special $\Lambda$-type three-level NIQBs show distinct energy extraction ratio variations due to excited-state lifetime differences. For short-lived excited-state nuclei (e.g., 154Gd), incomplete energy extraction occurs with reduced purity and mixed quantum states, originating from enhanced spontaneous emission during charging that disrupts inter-level energy transfer and limits extraction efficiency.
\begin{table}[htbp]
\renewcommand{\arraystretch}{2}
\setlength{\abovecaptionskip}{0pt}   % 标题与表上框，默认10pt
\setlength{\belowcaptionskip}{6pt}   % 标题与表主体，默认0pt
  \centering
  \caption{Characteristic parameters of a two-level NIQBs systems. $\varepsilon_{i}$ is the energy of state $|i\rangle$ ($i = 1,2$), with $\varepsilon_{1} = 0$ keV. The relativistic factor $\gamma$, determined by the one-photon resonance condition $\varepsilon_{2}-\varepsilon_{1}=\gamma(1+\beta)\hbar \omega_{p}$ ($\hbar \omega_{p}$ denotes the pump photon energy). $\tau_{2}$ is the half-life of isomer state. The multipolarity $\mu L_{ij}$, $\mu \in \{E,M\}$ and $\mathbb{B}_{ij}(\mu L_{ij})$ denote the reduced matrix element for the transition $|1\rangle \rightarrow |2\rangle$. $I_{1}^{\pi}$ and $I_{2}^{\pi}$ represent the angular momentum and parity of the ground and the isomeric states, respectively. The peak intensities of pump laser pulse also has been given \cite{NNDC2025}. }
  \scalebox{0.995}{
      \begin{tabular}{cccccccccccc}
      \hline\hline
       Nucleus &$\varepsilon_{2}$ (keV) &$\gamma$ &$\tau_{2}$ &$\mu L_{12}$ &$\mathbb{B}_{12}$ (wsu) &$I_{1}^{\pi}$ &$I_{2}^{\pi}$ &$\Omega_{p0}$ (1/s) &$I_{p}$ ($10^{22}$ W/cm$^{2}$) \\
       \hline
       $^{133}\text{Xe}$ &$233.22$ &$9.43$ &$2.198$ d &$E5$ &$1.1\times10^{3}$ &$3/2^{+}$ &$11/2^{-}$ &$6.0237\times10^{3}$ &$0.1531$  \\
       $^{193}\text{Ir}$ &$80.24$ &$3.31$ &$10.53$ d &$M4$ &$2.15$ &$3/2^{+}$ &$11/2^{-}$ &$16.4662$ &$2.4217\times10^{3}$  \\
       $^{117}\text{Sn}$ &$314.58$ &$12.70$ &$14$ d &$E5$ &$0.045$ &$1/2^{+}$ &$11/2^{-}$ &$98.1065$ &$1.0496\times10^{3}$  \\
       $^{127}\text{Te}$ &$88.23$ &$3.63$ &$106.1$ d &$M4$ &$3.6$ &$3/2^{+}$ &$11/2^{-}$ &$17.9322$ &$2.4760\times10^{3}$  \\
       $^{113}\text{Cd}$ &$263.54$ &$10.65$ &$14.1$ y &$E5$ &$0.0499$ &$1/2^{+}$ &$11/2^{-}$ &$40.2961$ &$4.3695\times10^{3}$  \\
       $^{93}\text{Nb}$ &$30.77$ &$1.44$ &$16.12$ y &$M4$ &$11.49$ &$9/2^{+}$ &$1/2^{-}$ &$0.4892$ &$4.0245\times10^{5}$  \\
      \hline\hline
      \end{tabular}
      }
  \label{table1}
\end{table}
\section{The parameters for nuclei}\label{section3}
The coefficient $\Omega_{0}$ associated with the nuclear characteristic parameters in Eq.~(2) of the main text can be explicitly written as \cite{RevModPhys.70.1003,PhysRevC.77.044602}
%\begin{eqnarray}
%\begin{split}
%\Omega_{p(s)}=\Omega_{0p(s)} \sqrt{I_{p(s)}} exp\left\{-\left[\frac{\gamma(1+\beta)(t-\tau_{p(s)})}{\sqrt{2}T_{p(s)}} \right]^{2} \right\},
%\end{split}
%\end{eqnarray}
\begin{eqnarray}
\Omega_{0p(s)}=\frac{4\sqrt{\pi}}{\hbar} \left[\frac{\gamma^{2}(1+\beta)^{2}I_{p(s)}(L_{ij}+1)(2I_{i}+1)\mathbb{B}_{ij}(\mu L_{ij})}{c \epsilon_{0}L_{ij}}\right]^{\frac{1}{2}} \frac{k^{L_{ij}-1}_{ij}}{(2L_{ij}+1)!!},
\end{eqnarray}
where the relativistic factor $\gamma=1/\sqrt{(1-\beta)}$, and $\beta=v / c$ with $c$ being the velocity of the light in vacuum. $I_{p(s)}$ is the effective peak intensity of pump or Stokes laser. $L_{ij}$ represents the multipolarity of nuclear $|i\rangle \rightarrow |j\rangle$ transition. $I_{i}$ is the nuclear spin of the level $|i\rangle$. $\mathbb{B}_{ij}(\mu L_{ij})$ is the reduced transition probability for nuclear electric or magnetic $|i\rangle \rightarrow |j\rangle$ transition, denoted by $\mu \in \{E, M\}$. The wave number $ k_{ij} $ satisfies the condition $\gamma (1+\beta) \hbar \omega_{p(s)}=c k_{ij}, \{i,j\}\in \{1,2,3\}$.
Tables~\ref{table1} and \ref{table2} present the nuclear characteristic parameters for the two-level and three-level systems constructed from the selected nuclei.

\begin{table}[htbp]
\renewcommand{\arraystretch}{2}
\setlength{\abovecaptionskip}{0pt}   % 标题与表上框，默认10pt
\setlength{\belowcaptionskip}{6pt}   % 标题与表主体，默认0pt
  \centering
  \caption{Characteristic parameters of a three-level nuclear system. $\varepsilon_{i}$ is the energy of state $|i\rangle$ ($i = 1,2,3$), with $\varepsilon_{1} = 0$ keV. The relativistic factor $\gamma$, determined by the one-photon resonance condition $\varepsilon_{2}-\varepsilon_{1}=\gamma(1+\beta)\hbar \omega_{p}$ ($\hbar \omega_{p}$ denotes the pump photon energy). $\tau_{2}$ is the half-life of state $|2\rangle$. $B_{2i}$ is the branching ratio of $|2\rangle \rightarrow |i\rangle$ ($i=1, 3$). The multipolarity $\mu L_{ij}$, $\mu \in\{E, M\}$ and $\mathbb{B}_{ij}(\mu L_{ij})$ denote the reduced matrix elements for the transitions $|i\rangle \rightarrow |j\rangle$ ($i, j=1, 2, 3$). $I_{1}^{\pi}$, $I_{2}^{\pi}$ and $I_{3}^{\pi}$ represent the angular momentum and parity of the ground, excited and isomeric states, respectively. $\tau_{p}$ and $\tau_{s}$ are the temporal peak positions of the pump and Stokes pulses. The peak intensities of pump and Stokes laser pulse also have been given \cite{NNDC2025}.  }
  \scalebox{0.66}{
      \begin{tabular}{cccccccccccccccccccc}
      \hline\hline
       Nucleus &$\varepsilon_{3}$ &$\varepsilon_{2}$ &$\gamma$ &$\tau_{2}$  &\multicolumn{2}{c}{Branching ratio} &\multicolumn{2}{c}{Multipolarity} &$\mathbb{B}_{12}$ &$\mathbb{B}_{23}$ &$I_{1}^{\pi}$ &$I_{2}^{\pi}$ &$I_{3}^{\pi}$ &$\tau_{p}$ &$\tau_{s}$ &$\Omega_{p0}$ &$\Omega_{s0}$ &$I_{p}$ &$I_{s}$ \\
       \textit{} &(keV) &(keV) &\textit{} &\textit{} \textit{} &$B_{21}$ &$B_{23}$ &$\mu L_{12}$ &$\mu L_{23}$ &(wsu) &(wsu) &\textit{} &\textit{} &\textit{} &(ps) &(ps) &(1/s) &(1/s) &($10^{22}$W/cm$^{2}$) &($10^{22}$W/cm$^{2}$)\\
       \hline
       $^{154}\text{Gd}$ &$123.00$ &$1241.00$ &$50.1$ &$1.54$ fs &$0.5167$ &$0.4752$ &$E1$ &$E1$ &$0.044$ &$0.049$ &$0^{+}$ &$1^{-}$ &$2^{+}$ &$0.016$ &$0.014$ &$2.8268\times$10$^{5}$ &$6.6704$ $\times$10$^{5}$ &$0.1421$ &$0.0255$ \\
       $^{63}\text{Ni}$ &$87.22$ &$1001.25$ &$40.38$ &$0.29$ ps &$0.0138$ &$0.9862$ &$M1$ &$E2$ &$2.3\times10^{-3}$ &$2.4$ &$1/2^{-}$ &$1/2^{-}$ &$5/2^{-}$ &$0.011$ &$0.009$ &$3.5878\times10^{2}$ &$4.0464\times10^{4}$ &$6.9960\times10^{4}$ &$5.50$  \\
       $^{229}\text{Th}$ &$0.008355$ &$29.19$ &$5.45$ &$82.2$ ps &$0.0936$ &$0.9250$ &$M1$ &$E2$ &$0.003$ &$44.9$ &$5/2^{+}$ &$5/2^{+}$ &$3/2^{+}$ &$1.35$ &$1.25$ &$2.7174\times10^{3}$ &$8.5621\times10^{3}$ &$0.5500$ &$0.0554$ \\
       $^{107}\text{Ag}$ &$93.125$ &$423.15$ &$17.08$ &$29.8$ ps &$0.4412$ &$0.5588$ &$E2$ &$E1$  &$43$ &$2.7\times 10^{-7}$ &$1/2^{-}$ &$5/2^{-}$ &$7/2^{+}$ &$0.0175$ &$0.0125$ &$2.7550\times$10$^{4}$ &$3.9208$ &$1.9320$ &$9.5040\times10^{7}$ \\
       $^{144}\text{Pr}$ &$59.03$ &$99.95$ &$4.09$ &$0.66$ ns &$0.094$ &$0.906$ &$E2$ &$E2$ &$56$ &$70$ &$0^{-}$ &$2^{-}$ &$3^{-}$ &$0.06$ &$0.04$ &$1.5097\times10^{3}$ &$1.8294\times10^{3}$ &$62.6150$ &$42.6580$ \\
       $^{103}\text{Rh}$ &$39.75$ &$357.39$ &$14.43$ &$73$ ps &$0.0137$ &$0.9863$ &$E2$ &$E1$ &$44$ &$7.9\times10^{-8}$ &$1/2^{-}$ &$5/2^{-}$ &$7/2^{+}$ &$0.018$ &$0.012$ &$1.9379\times10^{4}$ &$1.7686$ &$3.1972$ &$3.8392\times10^{8}$ \\
       $^{189}\text{Os}$ &$30.82$ &$216.67$ &$8.77$ &$0.41$ ns &$0.0199$ &$0.9801$ &$E2$ &$M1$ &$18.2$ &$1.05\times10^{-3}$ &$3/2^{-}$ &$7/2^{-}$ &$9/2^{-}$ &$0.035$ &$0.025$ &$9.7181\times10^{3}$ &$33.8532$ &$4.2704$ &$3.5196\times10^{5}$ \\
       $^{152}\text{Eu}$ &$45.60$ &$65.30$ &$2.73$ &$0.94$ $\mu s$ &$0.8007$ &$0.1993$ &$E2$ &$M1$ &$0.071$ &$7.9\times10^{-5}$ &$3^{-}$ &$1^{-}$ &$0^{-}$ &$0.096$ &$0.064$ &$63.0046$ &$4.1127$ &$2.1945\times10^{4}$ &$5.1500\times10^{6}$ \\
       $^{121}\text{Sn}$ &$6.31$ &$925.59$ &$37.33$ &$0.25$ ns &$0.9588$ &$0.0412$ &$E2$ &$M2$ &$0.081$ &$0.32$ &$3/2^{+}$ &$7/2^{+}$ &$11/2^{-}$ &$0.0113$ &$0.0087$ &$8.7790\times10^{3}$ &$3.0017\times10^{4}$ &$2.7254\times10^{2}$ &$23.3130$ \\
       $^{195}\text{Pt}$ &$259.08$ &$129.77$ &$5.28$ &$0.67$ ns &-- &-- &$E2$ &$M4$ &$8.9$ &$0.00107$ &$1/2^{-}$ &$5/2^{-}$ &$3/2^{+}$ &$0.11$ &$0.09$ &$1.7579\times10^{3}$ &$3.0979$ &$66.7920$ &$2.1508\times10^{7}$ \\
       $^{129}\text{Xe}$ &$236.14$ &$39.58$ &$1.75$ &$0.97$ ns &-- &-- &$E2$ &$M4$ &$9$ &$1.777$ &$1/2^{+}$ &$3/2^{+}$ &$11/2^{+}$ &$0.18$ &$0.12$ &$1.2442\times10^{2}$ &$63.5405$ &$8.2404\times10^{4}$ &$3.1596\times10^{5}$ \\
       $^{121}\text{Te}$ &$293.97$ &$212.19$ &$8.59$ &$0.062$ ns &-- &-- &$E2$ &$M4$ &$27$ &$3.18$ &$1/2^{+}$ &$3/2^{+}$ &$11/2^{-}$ &$0.055$ &$0.045$ &$5.9632\times10^{3}$ &$30.1499$ &$25.8990$ &$1.0132\times10^{6}$ \\
       $^{119}\text{Sn}$ &$89.53$ &$23.87$ &$1.22$ &$18.03$ ns &-- &-- &$E2$ &$M4$ &$0.7$ &$5.20$ &$1/2^{+}$ &$3/2^{+}$ &$11/2^{-}$ &$0.29$ &$0.21$ &$11.9266$ &$2.1912$ &$5.6945\times10^{4}$ &$1.6863\times10^{6}$ \\
       $^{108}\text{Ag}$ &$109.47$ &$79.14$ &$3.27$ &$1.2$ ns &-- &-- &$E1$ &$M4$ &$0.0005$ &$5.3\times10^{3}$ &$1^{+}$ &$2^{-}$ &$6^{+}$ &$0.165$ &$0.135$ &$22.3824$ &$22.5334$ &$2.5552\times10^{5}$ &$2.5212\times10^{5}$ \\
       $^{186}\text{Re}$ &$148.20$ &$99.36$ &$4.07$ &$25.5$ ns &-- &-- &$E2$ &$E5$ &$0.391$ &$0.069$ &$1^{-}$ &$3^{-}$ &$8^{+}$ &$0.165$ &$0.135$ &$2.5654\times10^{2}$ &$0.0765$ &$4.9066\times10^{4}$ &$5.5198\times10^{11}$ \\
      \hline\hline
      \end{tabular}
      }
  \label{table2}
\end{table}

\section{Performance metrics of NIQBs}\label{section4}
Key performance parameters for different nuclear species used as storage cells in two-level, three-level $\Lambda$-type and ladder-type configurations are summarized in Tables~\ref{table3} and \ref{table4}. We also calculated the stored energy and average charging power of the three-level atomic QBs corresponding to these typical nuclear species. The intrinsic properties, laser parameters, and QB performance are summarized in Table~\ref{table5}.
\begin{table}[htbp]
\renewcommand{\arraystretch}{1.8}
\setlength{\abovecaptionskip}{0pt}   % 标题与表上框，默认10pt
\setlength{\belowcaptionskip}{6pt}   % 标题与表主体，默认0pt
  \centering
  \caption{The intrinsic properties of typical nuclei, laser parameters and corresponding performance metrics of the two-level NIQBs.
  %The lifetime of the different nuclear isomers are $\tau_{2}$. $\varepsilon_{i} (i=1,2,3)$ are the energies of the ground-, isomeric- and excited-states of nuclei.
  The half-lives (proportional to the lifetimes) and energies of the different nuclear ground and isomeric states are given by $\tau_{i}$ and $\varepsilon_{i}$~($i=1, 2$, and $\varepsilon_{1}=0$~kev). The decay mode refers to that of the ground state. $\Omega_{\text{p}}^\text{max}$, $t_{sta}$, $t_{tot}$, $E^{\text{sta}}$, $P^{\text{max}}$ and $W^{\text{sta}}$ represent the peak pulse intensity, stable stored energy time, total evolution time, stable stored energy, peak charging power and stable ergotropy, respectively.}% ``$-$" indicates that the decay is forbidden.}
  \scalebox{0.90}{
      \begin{tabular}{ccccccccccc}
       %\toprule
      \hline\hline
       Nucleus &\multicolumn{4}{c}{intrinsic properties} &\multicolumn{2}{c}{laser parameters} &\multicolumn{3}{c}{charging performance} \\
       \cmidrule{2-8}
       \textit{} &Decay mode &$\tau_{1}$ &$\tau_{2}$ &$\varepsilon_{2}$ (keV) &$\Omega_{p}^{\text{max}}$ ($1/s$) &$t_{\text{sta}}~(t_{\text{tot}})$ ($ps$) &$E^{\text{sta}}$ (keV) &$P^{\text{max}}$ (W) &$W^{\text{sta}}$ (keV)  \\
       \hline
        $^{133}\text{Xe}$ &$\beta$ &$5.2475$ d &$2.198$ d &$233.22$ &$2.36\times10^{14}$ &$0.08~(0.10)$ &$233.22$ &$0.64$ &$233.22$  \\
        $^{193}\text{Ir}$ &\text{stable} &-- &$10.53$ d &$80.24$ &$8.10\times10^{13}$ &$0.09~(0.10)$ &$80.24$ &$0.18$ &$80.24$  \\
        $^{117}\text{Sn}$ &\text{stable} &-- &$14$ d &$314.58$ &$3.18\times10^{14}$ &$0.07~(0.10)$ &$314.58$ &$0.89$ &$314.58$  \\
        $^{127}\text{Te}$ &$\beta$ &$9.35$ h &$106.1$ d &$88.23$ &$8.92\times10^{13}$ &$0.09~(0.10)$ &$88.23$ &$0.21$ &$88.23$  \\
        $^{113}\text{Cd}$ &$\beta$ &$8.04\times10^{15}$ &$14.1$ y &$263.54$ &$2.66\times10^{14}$ &$0.07~(0.10)$ &$263.54$ &$0.73$ &$263.54$  \\
        $^{93}\text{Nb}$ &\text{stable} &-- &$16.12$ y &$30.77$ &$3.10\times10^{13}$ &$0.26~(0.30)$ &$30.77$ &$0.02$ &$30.77$ \\
       \hline\hline
        %\bottomrule
      \end{tabular}
      }
  \label{table3}
\end{table}

\begin{table}[htbp]
\renewcommand{\arraystretch}{1.8}
\setlength{\abovecaptionskip}{0pt}   % 标题与表上框，默认10pt
\setlength{\belowcaptionskip}{6pt}   % 标题与表主体，默认0pt
  \centering
  \caption{The intrinsic properties of typical nuclei, laser parameters and corresponding performance metrics of the three-level NIQBs.
  %The lifetime of the different nuclear isomers are $\tau_{2}$. $\varepsilon_{i} (i=1,2,3)$ are the energies of the ground-, isomeric- and excited-states of nuclei.
  The half-lives (proportional to the lifetimes) and energies of the different nuclear ground, isomeric, and excited states are given by $\tau_{i}$ and $\varepsilon_{i}~(i=1, 2, 3$, and $\varepsilon_{1}=0$~kev). The decay mode refers to that of the ground state. $\Omega_{\text{p/s}}^\text{max}$, $A$, $t_{sta}$, $t_{tot}$, $E^{\text{sta}}$, $E^{\text{max}}$, $P^{\text{max}}$, $W^{\text{sta}}$ and $W^{\text{max}}$ represent the peak pulse intensity, pulse area, stable stored energy time, total evolution time, stable stored energy, maximum stored energy, peak charging power, stable ergotropy and maximum ergotropy, respectively.}% ``$-$" indicates that the decay is forbidden.}
  \scalebox{0.72}{
      \begin{tabular}{ccccccccccccccc}
       %\toprule
      \hline\hline
       Nucleus &\multicolumn{6}{c}{intrinsic properties} &\multicolumn{3}{c}{laser parameters} &\multicolumn{3}{c}{charging performance} &\multicolumn{1}{c}{scheme} \\
       \cmidrule{2-8}
       \textit{} &Decay mode &$\tau_{1}$ &$\tau_{2}$ &$\varepsilon_{2}$ (keV) &$\tau_{3}$ &$\varepsilon_{3}$ (keV) &$\Omega_{p/s}^{\text{max}}$ ($1/s$) &$A$ &$t_{\text{sta}}~(t_{\text{tot}})$ ($ps$) &$E^{\text{sta}}$ ($E^{\text{max}}$) (keV) &$P^{\text{max}}$ (W) &$W^{\text{sta}}$ ($W^{\text{max}}$) (keV) &\textit{} \\
       \hline
        $^{154}\text{Gd}$ &\text{stable} &-- &$1.54$ fs &$1241.00$ &$1.184$ ns &$123.00$ &$1.07\times10^{16}$ &$26.66$ &$0.02~(0.03)$ &$121.69~(177.01)$ &$1.85$ &$120.47~(171.85)$ &$\Lambda$-type \\
        $^{63}\text{Ni}$ &$\beta$ &$100.8$ y &$0.29$ ps &$1001.25$ &$1.69$ $\mu$s &$87.22$ &$9.49\times10^{15}$ &$29.46$ &$0.01~(0.02)$ &$87.23~(89.11)$ &$1.31$ &$87.20~(89.08)$ &$\Lambda$-type \\
        $^{229}\text{Th}$ &$\alpha$ &$7907$ y &$82.2$ ps &$29.19$ &$7$ $\mu$s &$8.355\times10^{-3}$ &$3.2 \times 10^{15}$ &$62.83$ &$1.73~(2.60)$ &$8.355\times10^{-3}~(0.22)$ &$2.5 \times 10^{-5}$ &$8.354\times10^{-3}~(0.20)$ &$\Lambda$-type \\
        $^{107}\text{Ag}$ &\text{stable} &-- &$29.8$ ps &$423.15$ &$44.3$ s &$93.13$ &$3.82\times10^{15}$ &$28.10$ &$0.02~(0.03)$ &$93.13~(93.59)$ &$0.84$ &$93.13~(93.59)$ &$\Lambda$-type \\
        $^{144}\text{Pr}$ &$\beta$ &$17.28$ min &$0.66$ ns &$99.95$ &$7.2$ min &$59.03$ &$1.19\times10^{15}$ &$37.15$ &$0.08 ~(0.10)$ &$59.03~(59.03)$ &$0.15$ &$59.03~(59.03)$ &$\Lambda$-type \\
        $^{103}\text{Rh}$ &\text{stable} &-- &$73$ ps &$357.40$ &$56.114$ min &$39.75$ &$3.47\times10^{15}$ &$30.12$ &$0.02 ~(0.03)$ &$39.74~(40.52)$ &$0.38$ &$39.74~(40.52)$ &$\Lambda$-type \\
        $^{189}\text{Os}$ &\text{stable} &-- &$0.41$ ns &$216.67$ &$5.81$ h &$30.82$ &$2.01\times10^{15}$ &$28.79$ &$0.04 ~(0.06)$ &$30.81~(31.25)$ &$0.14$ &$30.81~(31.25)$ &$\Lambda$-type \\
        $^{152}\text{Eu}$ &$\beta$ &$13.52$ y &$0.94$ $\mu$s &$65.30$ &$9.3$ h &$45.60$ &$9.33\times10^{14}$ &$44.41$ &$0.13 ~(0.16)$ &$45.60~(45.60)$ &$0.07$ &$45.60~(45.60)$ &$\Lambda$-type \\
        $^{121}\text{Sn}$ &$\beta$ &$27.03$ h &$0.25$ ns &$925.59$ &$43.9$ y &$6.31$ &$8.37\times10^{15}$ &$48.67$ &$0.01~(0.02)$ &$6.37~(18.22)$ &$0.29$ &$6.37~(18.22)$ &$\Lambda$-type \\
        $^{195}\text{Pt}$ &\text{stable} &-- &$0.67$ ns &$129.77$ &$4.01$ d &$259.08$ &$1.44\times10^{15}$ &$34.41$ &$0.16~(0.20)$ &$258.05~(258.95)$ &$0.38$ &$258.05~(258.95)$ &$\text{ladder-type}$ \\
        $^{129}\text{Xe}$ &\text{stable} &-- &$0.97$ ns &$39.58$ &$8.88$ d &$236.14$ &$3.57\times10^{15}$ &$280.97$ &0.24~(0.30) &$236.14~(236.14)$ &$0.21$ &$236.14~(236.14)$ &$\text{ladder-type}$ \\
        $^{121}\text{Te}$ &$\beta$ &$19.17$ d &$0.062$ ns &$212.19$ &$164.2$ d &$293.97$ &$3.03\times10^{15}$ &$44.43$ &$0.07(0.10)$ &$293.90~(293.91)$ &$0.82$ &$293.90~(293.91)$ &$\text{ladder-type}$ \\
        $^{119}\text{Sn}$ &\text{stable} &-- &$18.03$ ns &$23.87$ &$293.1$ d &$89.53$ &$2.85\times10^{14}$ &$37.17$ &$0.41~(0.50)$ &$89.53~(89.53)$ &$0.05$ &$89.53~(89.53)$ &$\text{ladder-type}$  \\
        $^{108}\text{Ag}$ &$\beta$ &$2.382$ min &$1.2$ ns &$79.14$ &$438$ y &$109.47$ &$1.13\times10^{15}$ &$44.43$ &$0.25~(0.30)$ &$109.45~(109.47)$ &$0.10$ &$109.45~(109.47)$ &$\text{ladder-type}$ \\
        $^{186}\text{Re}$ &$\beta$ &$3.7185$ d &$25.5$ ns &$99.36$ &$2\times10^{5}$ y &$148.20$ &$5.68\times10^{15}$ &$177.71$ &$0.20~(0.30)$ &$148.20~(148.20)$ &$0.15$ &$148.20~(148.20)$ &$\text{ladder-type}$ \\
        %$^{178}\text{Hf}$ &$\gamma$ &-- &$2136.52$ &$31$ y &$2446.09$ &$68$ $\mu$s &$2572.40$ &$5.21\times10^{15}$ &$34.41$ &$0.02$ &$309.35$ &$2.83$ &$309.35$ &$99.99\%$  \\
       \hline\hline
        %\bottomrule
      \end{tabular}
      }
  \label{table4}
\end{table}

\begin{table}[htbp]
\renewcommand{\arraystretch}{1.8}
\setlength{\abovecaptionskip}{0pt}   % 标题与表上框，默认10pt
\setlength{\belowcaptionskip}{6pt}   % 标题与表主体，默认0pt
  \centering
  \caption{The intrinsic properties of atom, evolution time and corresponding performance metrics of the three-level atomic QB. The energies of the different atom ground, first excited, and second excited states are given by $\varepsilon_{i}$~($i=1, 2, 3$, and $\varepsilon_{1}=0$ eV). Peak pulse intensity $\Omega_{\text{p/s}}^\text{max} = 5\times10^{9}$. Temporal peak positions of the pump and Stokes pulses: $\tau_p = 5.46~\text{ns}$ and $\tau_s = 3.54~\text{ns}$. The total evolution time is set to $t_{tot} = 9.00~\text{ns}$. $t_{sta}$, $E^{\text{sta}}$ and $P^{\text{max}}$ represent stable stored energy time, stable stored energy and peak charging power, respectively.}
   \scalebox{0.9}{
      \begin{tabular}{cccccccccccccc}
       %\toprule
      \hline\hline
       Atoms &\multicolumn{2}{c}{intrinsic properties} &\multicolumn{1}{c}{evolution time} &\multicolumn{2}{c}{charging performance} &Atoms &\multicolumn{2}{c}{intrinsic properties} &\multicolumn{1}{c}{evolution time} &\multicolumn{2}{c}{charging performance} \\
       \cmidrule{2-6} \cmidrule{8-12}
       \textit{} &$\varepsilon_{2}$ (eV) &$\varepsilon_{3}$ (eV) &$t_{\text{sta}}$ ($ns$) &$E^{\text{sta}}$ (eV) &$P^{\text{max}}$ (W) &\textit{} &$\varepsilon_{2}$ (eV) &$\varepsilon_{3}$ (eV) &$t_{\text{sta}}$ ($ns$) &$E^{\text{sta}}$ (eV) &$P^{\text{max}}$ (W)  \\
       \hline
        $\text{Gd}$ &$0.03$ &$0.07$ &$6.76$ &$0.07$ &$1.87\times10^{-12}$ &$\text{Pt}$ &$0.81$ &$1.26$ &$6.97$ &$1.26$ &$3.56\times10^{-11}$ \\
        $\text{Ni}$ &$0.17$ &$0.27$ &$6.88$ &$0.27$ &$7.78\times10^{-12}$ &$\text{Eu}$ &$1.97$ &$1.98$ &$6.94$ &$1.98$ &$5.61\times10^{-11}$ \\
        $\text{Rh}$ &$0.19$ &$0.32$ &$6.90$ &$0.32$ &$9.11\times10^{-12}$ &$\text{Re}$ &$3.58$ &$3.58$ &$6.97$ &$3.58$ &$1.02\times10^{-10}$ \\
        $\text{Pr}$ &$0.17$ &$0.35$ &$6.91$ &$0.35$ &$9.98\times10^{-12}$ &$\text{Ag}$ &$3.66$ &$3.78$ &$6.97$ &$3.78$ &$1.07\times10^{-10}$ \\
        $\text{Sn}$ &$0.21$ &$0.42$ &$6.93$ &$0.42$ &$1.20\times10^{-11}$ &$\text{Te}$ &$0.59$ &$5.78$ &$7.01$ &$5.78$ &$1.63\times10^{-10}$ \\
        $\text{Os}$ &$0.34$ &$0.52$ &$6.92$ &$0.52$ &$1.46\times10^{-11}$ &$\text{Xe}$ &$8.44$ &$9.45$ &$7.00$ &$9.45$ &$2.68\times10^{-10}$ \\
        $\text{Th}$ &$0.36$ &$0.62$ &$6.94$ &$0.62$ &$1.74\times10^{-11}$ \\

       \hline\hline
        %\bottomrule
      \end{tabular}
      }
  \label{table5}
\end{table}

\end{document}